\documentclass[twocolumn,tighten,times]{aastex62}

\usepackage{amssymb,amsmath}
\usepackage[inline]{enumitem} 
\usepackage{soul}
\usepackage{natbib}
\usepackage{color}
\newcommand{\kms}{km~s$^{-1}\,$}

\newcommand{\HII}{H {\small{II}} }

\usepackage{longtable}
\usepackage{booktabs}

\shorttitle{The Formation of the AFGL 333-Ridge}
\shortauthors{Liang et al.}

\begin{document}

\title{ {\bf A Multi-wavelength Study on the Formation of AFGL 333-Ridge} }

\correspondingauthor{Jin-Long Xu}
\email{xujl@bao.ac.cn}

\author{Xiaolian Liang}
\affil{National Astronomical Observatories, Chinese Academy of Sciences, Beijing 100101, People's Republic of China}
\affil{University of Chinese Academy of Sciences, Beijing 100049, People's Republic of China}

\author{Jin-Long Xu}
\affil{National Astronomical Observatories, Chinese Academy of Sciences, Beijing 100101, People's Republic of China}

\author{Ye Xu}
\affiliation{Purple Mountain Observatory, Chinese Academy of Sciences, Nanjing 210008, People's Republic of China}

\author{Jun-Jie Wang}
\affil{National Astronomical Observatories, Chinese Academy of Sciences, Beijing 100101, People's Republic of China}

\begin{abstract}
We presented a multi-wavelength study of AFGL 333-Ridge. The molecular line data reveals that the AFGL 333-Ridge has two independent velocity components at $-$50.5 km s$^{-1}$ and $-$48.0 km s$^{-1}$. In the Position-Velocity diagram, the bridge feature connects with two parts that are spatially correlated but separated in velocity. These observational evidences support the scenario that the two velocity components have collided and merged into one molecular cloud. The majority of Class I YSOs are distributed within the collision region, suggesting that the cloud-cloud collision has induced the YSOs formation in the ridge. Using the radio recombination line (RRL) data obtained by the Five-hundred-meter Aperture Spherical radio Telescope (FAST), the RRL velocities of three \HII regions are consistent with that of the AFGL 333-Ridge. By comparing the three \HII regions' dynamical ages with the collision timescale of the two components, we conclude that the influence of the three \HII regions may not drive the two clouds to merge. The formation of the AFGL 333-Ridge is probably due to the expansion of the giant \HII region W4.
\end{abstract}


\section{Introduction}
Dust continuum Galactic Plane surveys have detected abundant filaments \citep{2006AJ....131.1163S,2010A&A...518L.100M}. Since the star-forming cores were found distributed along the filaments \citep{2010A&A...518L.102A,2015PASA...32....7A,2017CRGeo.349..187A}, the filamentary structure may play an important role in star formation.  But how the filamentary cloud form is still unclear. In Jeans-unstable clouds, gravity cause large-scale collapse first along the shortest axis of these clouds, followed by filaments formation \citep{2017CRGeo.349..187A}. However, in gravitationally unbound  clouds, the formation of many filaments cannot be explained by gravity. Recently, there is growing research that filaments may form due to turbulence' dissipation on a large scale \citep{2011A&A...529L...6A,2012A&A...541A..63P,2019ApJ...878..157X}. These large-scale supersonic flows create the locally planar shock,  which can compress interstellar clouds and then create a universal filamentary structure \citep{2014prpl.conf...27A}. When considering the combined effect of gravity and turbulence, it is a plausible mechanism to explain the filamentary web structure and hub-filament structure \citep{2011A&A...529L...6A,2012A&A...541A..63P}. Additionally, according to the MHD model \citep{2010ApJ...720L..26L}, the magnetic field can strongly affect the turbulent flow, and the masses of filaments will accumulate along the field lines. Since the magnetic field could support the stability of filament, it can make the filaments survive longer \citep{2017MmSAI..88..513H}. At present, global gravity, large-scale turbulence, and magnetic field may play the dominant role in forming filaments \citep{2011A&A...529L...6A}. ``Ridge'' usually refers to the high-density filaments, whose column  density are greater than  $10^{23}$ cm$^{-2}$ \citep{2011A&A...533A..94H, 2012A&A...543L...3H}. Further, studying the properties of the ridge is also helpful for us to reveal the formation of the filaments. 

\begin{figure*}
    \centering
    \includegraphics[width=0.85\textwidth]{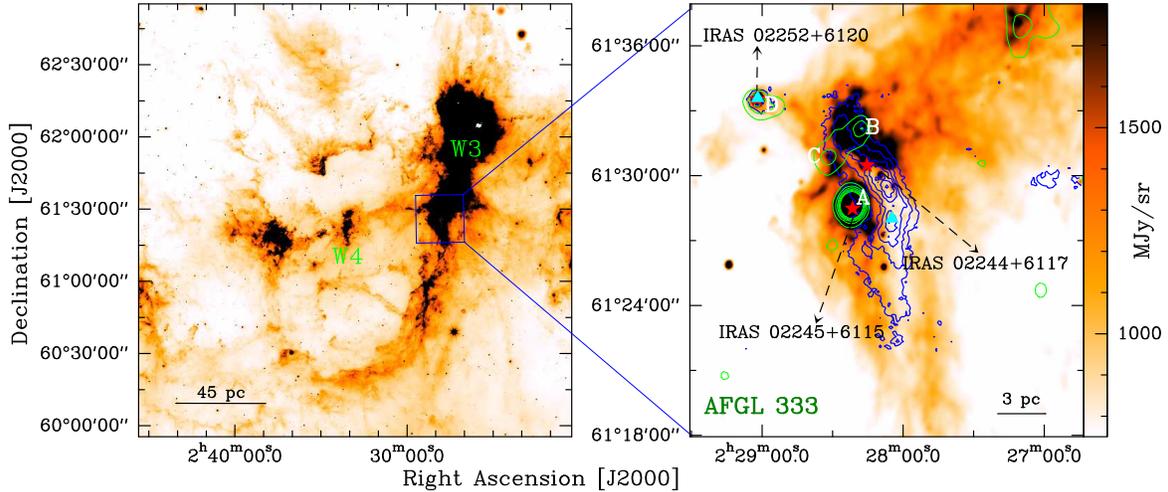}
    \caption{Left panel: the WISE 12 $\mu$m image (color) of the W3 GMC and W4 giant \HII region.
The blue rectangle shows the area of AFGL 333-Ridge. Right panel:  the WISE 12 $\mu$m zoomed image  (color) of AFGL 333-Ridge overlaid with the BGPS 1.1 mm continuum emission (blue contours) and NVSS 1.4 GHz radio continuum emission (green contours). To show better the background emission for \HII A,  the dense contours of the radio continuum emission are cut off. The blue 1.1 mm emission contours are shown by the levels of 0.05 Jy/beam $\times$ (3, 8, 13, 18, 23, 28, 33) and the NVSS contours start from 3.5 mJy/beam (5$\sigma$) in steps of 5$\sigma$. The red star and cyan triangle symbols indicate the positions of three IRAS sources and two H$_2$O masers, respectively. The white letter A, B, C, and D illustrate the location of four \HII regions.
    }
    \label{fig:w34}
\end{figure*}

Giant molecular cloud (GMC) W3 contains three active star-forming regions, W3 Main, W3(OH), and AFGL 333-Ridge \citep{1985ApJ...297..662T}. These three regions are similar in size and mass, but they show very different star-forming activities \citep{2015ApJ...809...81R}. AFGL 333-Ridge is not well studied as W3 main and W3(OH), and its star formation efficiency is lower than the other two regions \citep{2013ApJ...766...85R}. The AFGL 333-Ridge shows a filamentary structure \citep{2013ApJ...766...85R}, which is associated with three IRAS sources (IRAS 02244+6117, IRAS 02245+6115, and IRAS 02252+6120), and two H$_{2}$O masers \citep{2017PASJ...69...16N}. A CO bipolar outflow has been identified in the southern H$_{2}$O maser position \citep{2017PASJ...69...16N}. The parallax distance of W3 (OH) was obtained to be 2.0 kpc \citep{2006Sci...311...54X}. We will assume a common distance of 2.0 kpc for the AFGL 333-Ridge. Moreover, the AFGL 333-Ridge is located on the western edge of the giant \HII region W4 \citep{1978ApJ...226L..39L}. \cite{2017PASJ...69...16N} indicated that the overall structure of the AFGL 333-Ridge might have been formed by the expanding bubble of W4. But there is no evidence to support that the star-forming activities in this region are impacted by the feedback of the giant \HII region W4 \citep{2016ApJ...822...49J,2017PASJ...69...16N}.

This paper is organized as follows. In Section 2, we will introduce the multi-wavelength observations. Section 3 shows the identification of the AFGL 333-Ridge filament, the physical properties of  dust and gas, and the distribution of embedded YSOs in the  AFGL 333-Ridge. Section 4 is devoted to explain the possible reason of the AFGL 333-Ridge formation and local star formation scenario in this region. Finally, we will give a summary in Section 5.

\section{Observation and data processing}
\label{sect:data}
\subsection{Molecular line data}
Toward the AFGL 333-Ridge, we mapped a $15^{\prime}\times 15^{\prime}$ region centered at R.A.= 02:28:07 and Dec.= 61:27:56 in the transitions of $^{12}$CO $J$=1-0 (115.271 GHz), $^{13}$CO $J$=1-0 (110.201 GHz), C$^{18}$O $J$=1-0 (109.782 GHz), HCO$^+$ $J$=1-0 (89.188 GHz) and N$_2$H$^+$ $J$=1-0 (93.173 GHz) molecular lines using the PMO 13.7m radio telescope at De Ling Ha in western China, during January 2014 and May 2020. The 3$\times$3 beam array receiver system in single-side band (SSB) mode was used as front end. The back end is a Fast Fourier Transform Spectrometer (FFTS) of 16384 channels with a bandwidth of 1 GHz, corresponding to a velocity resolution of 0.16 km s$^{-1}$ for $^{12}$CO $J$=1-0, 0.17 km s$^{-1}$ for $^{13}$CO $J$=1-0 and C$^{18}$O $J$=1-0, and 0.20 km s$^{-1}$ for HCO$^+$ $J$=1-0 and N$_2$H$^+$ $J$=1-0. The half-power beam width (HPBW) was 53$^{\prime\prime}$ at 115 GHz and the main beam efficiency was 0.5. The pointing accuracy of the telescope was better than 5$^{\prime\prime}$. Mapping observations use the on-the-fly mode with a constant integration time of 14 seconds at each point and with a $0.5^{\prime}\times0.5^{\prime}$ grid. The final data were recorded in brightness temperature scale of $T_{\rm mb}$ (K) and were reduced using the GILDAS/CLASS\footnote{http://www.iram.fr/IRAMFR/GILDAS/} package.

\subsection{Radio recombination line data}
Moreover, we observed four \HII regions adjacent to the AFGL 333-Ridge filament in the C166$\alpha$ (1425.44447 MHz) and H166$\alpha$ (1424.73359 MHz) radio recombination lines (RRLs) using the Five-hundred-meter Aperture Spherical radio Telescope (FAST), during  March 2020. which is located in Guizhou, China. The aperture of the telescope is 500 m and the effective aperture is about 300 m. The half-power beam width (HPBW) was 2.9$^{\prime}$ and the velocity resolution is 0.1 km s$^{-1}$ for the digital backend at 1.4 GHz. Single pointing observation of the four \HII regions was taken with the wide mode of spectral backend. This mode has 1048576 channels in 500 MHz bandwidth. During this observation, system temperature was around 36 K. For the four HII regions, each observed position was integrated for 30 minutes. Reduced root mean square (rms) of the final spectrum is shown in Table 1. \cite{2019SCPMA..6259502J} gave more details of the Five-hundred-meter Aperture Spherical radio Telescope.

\subsection{Archival data}
\label{sect:archive}
 In order to obtain the dust temperature and column density, we use the observation data of {\it {\it Herschel}} Infrared Galactic Plane survey \citep{2010A&A...518L.100M}. The Hi-GAL survey is a {\it {\it Herschel}} key project that covers a longitude range of $\mid l \mid \leq$ 60.0 and a latitude range of $\mid b \mid \leq$ 1.0 map \citep{2010PASP..122..314M}. The {\it {\it Herschel}} data include PACS 70 and 160 $\mu$m \citep{2010A&A...518L...2P} and SPIRE 250, 350 and 500 $\mu$m \citep{2010A&A...518L...3G}. The beam sizes of the {\it {\it Herschel}} maps are 8.4$^{\prime\prime}$, 13.5$^{\prime\prime}$, 18.2$^{\prime\prime}$, 24.9$^{\prime\prime}$ and 36.3$^{\prime\prime}$ at 70, 160, 250, 350, and 500 $\mu$m, respectively.
 
To trace the dense gas of AFGL 333-Ridge, we use 1.1 mm continuum data of the Bolocam Galactic Plane Survey (BGPS), which covers a longitude range of -10.5 $^{\circ} \leq l \leq$ 90.5$^{\circ}$ and a latitude range of $\mid b \mid \leq$ 0.5 \citep{2011ApJS..192....4A}.

To trace the \HII region adjacent to AFGL 333-Ridge, we use the Wide-field Infrared Survey Explorer (WISE) observed the whole sky in 12 $\mu$m infrared bands whose resolution is 6.5$^{\prime \prime}$ \citep{2010AJ....140.1868W}. And the 4.5 $\mu$m emission data was obtained from {\it{\it Spitzer}}-GLIMPSE whose resolution is 2$^{\prime \prime}$. To trace the ionizing gas of \HII region, we also used the NARO VLA Sky Survey (NVSS) 1.4 GHz (21 cm) continuum emission data \citep{1998AJ....115.1693C}.

\section{Results}
\label{sec:results}
\subsection{Infrared and continuum emission}
\label{sec:ifc}
Figure~\ref{fig:w34} (left panel) shows a WISE 12 $\mu$m image of GMC W3 and giant \HII W4. The blue rectangle indicates the position and area of the AFGL 333-Ridge. The WISE 12 $\mu$m emission is similar to the Spitzer IRAC 8.0 $\mu$m emission, which can be used to trace PDR (Photo-Dissociation Region), and delineates \HII region boundaries \citep{2008ARA&A..46..289T,2009A&A...494..987P}. Figure~\ref{fig:w34} (right panel) presents an enlarged WISE 12 $\mu$m emission map of the AFGL 333-Ridge, and those blue (the 1.1 mm continuum emission) and green (the NVSS 1.4 GHz continuum emission) contours are superimposed on the WISE image. The 1.1 mm continuum emission originates from cold dust. The cold dust emission of the AFGL 333-Ridge shows an S-like structure elongated from north to south. \cite{2017PASJ...69...16N} has indicated that the AFGL 333-Ridge is associated with three IRAS sources (IRAS 02244+6117, IRAS 02245+6115, and IRAS 02252+6120) which highlighted by red star symbols and two H$_2$O masers represented by cyan triangle marks in Figure~\ref{fig:w34}. Moreover, the 1.4 GHz radio continuum emission is mainly from free-free emission, which can trace the ionized gas of \HII regions. From the ionized gas emission, as shown in green contours in Figure~\ref{fig:w34} (right panel), we identify four \HII regions, named \HII regions A, B, C, and D. In space, the \HII region A may be related to IRAS 02245+6115 and a compact \HII region G134.2+0.8 \citep{1982AJ.....87..685H}. The 12 $\mu$m observations are a good tracer of PAH (Polycyclic Aromatic Hydrocarbons) emission. Figure~\ref{fig:w34} (right panel), shows that each \HII region is associated with some strong PAH emission in black. Significantly, the PAH emission surrounding the \HII region B shows an arc-like structure with an opening towards the northwest. Near \HII regions A and B, the AFGL 333-Ridge also shows an arc-like structure. In the northeast of AFGL 333-Ridge, a compact dust core is associated with \HII region D, IRAS 02252+6120, and bright-rimmed cloud SFO 05 \citep{2013ApJ...773..132F}. 

\begin{figure}
    \centering
    \includegraphics[width=0.45\textwidth]{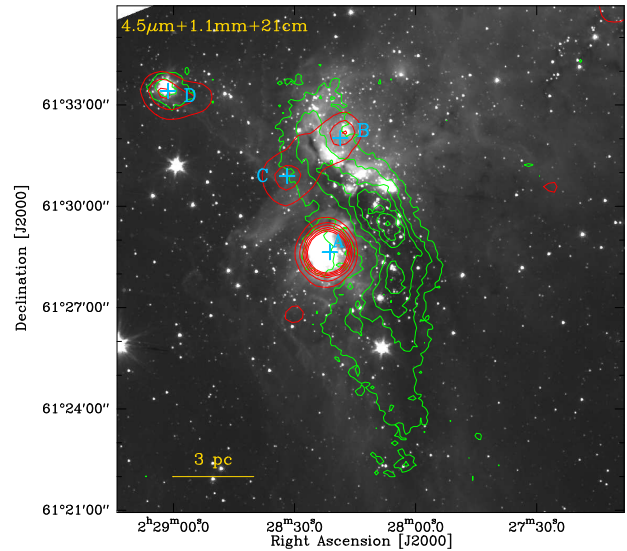}
    \caption{ 
       Overlay of the NVSS 1.4 GHz radio continuum emission contours (in red) and the BGPS 1.1 mm dust emission contours (in green) on the 4.5 $\mu$m infrared emission image. The blue plus symbols indicate the central position of the four \HII regions.}
    \label{fig:arc}
\end{figure}

Figure~\ref{fig:arc} displays the Spitzer-IRAC 4.5 $\mu$m emission (gray scale) overlaid with the 1.4 GHz continuum emission image (red contours) and the BGPS 1.1 mm dust emission contours (green contours). The Spitzer-IRAC 4.5 $\mu$m band contains both H$_{2}$ ($\rm \upsilon$= 0--0, S(9, 10, 11)) lines and CO ($\rm \upsilon$= 1--0) line, as shown in Figure 1 of \citet{2006AAS...20919204R}. \cite{2008AJ....136.2391C} noticed that all of these lines might be excited by shocks. Hence, the 4.5 $\mu$m  emission can be used to trace shocked molecular gas. In Figure~\ref{fig:arc}, for the \HII region B, the 4.5 $\mu$m  emission (gray scale) shows an arc-like structure similar to that of the PAH emission. The arc-like 4.5 $\mu$m emission may indicate the gas shocked by the \HII region B. Besides, we can see that the bright 4.5 $\mu$m  emission is concentrated on \HII regions D, indicating that the \HII region D is associated with an active massive star formation region.

\begin{figure*}
    \centering
    \includegraphics[width=0.9\textwidth]{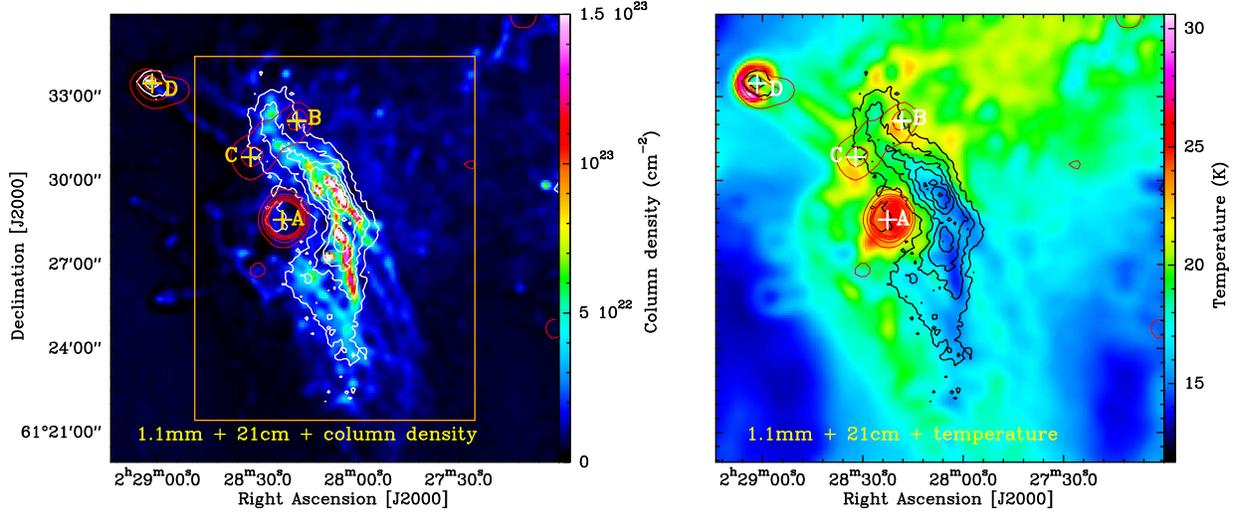}
    \caption{ Left panel: the BGPS 1.1 mm continuum emission (white contours) and the NVSS 1.4GHz continuum emission (red contours) superposed on the Herschel column density map. The orange box shows the target region for spectra analysis. Right panel: the BGPS 1.1 mm continuum emission (black contours) and the NVSS 1.4GHz continuum emission (brown contours) overlaid on the Herschel dust temperature map. The orange and white plus symbols indicate the central position of the four \HII regions.}
    \label{fig:sed}
\end{figure*}

\begin{figure}
    \centering
    \includegraphics[width=0.45\textwidth]{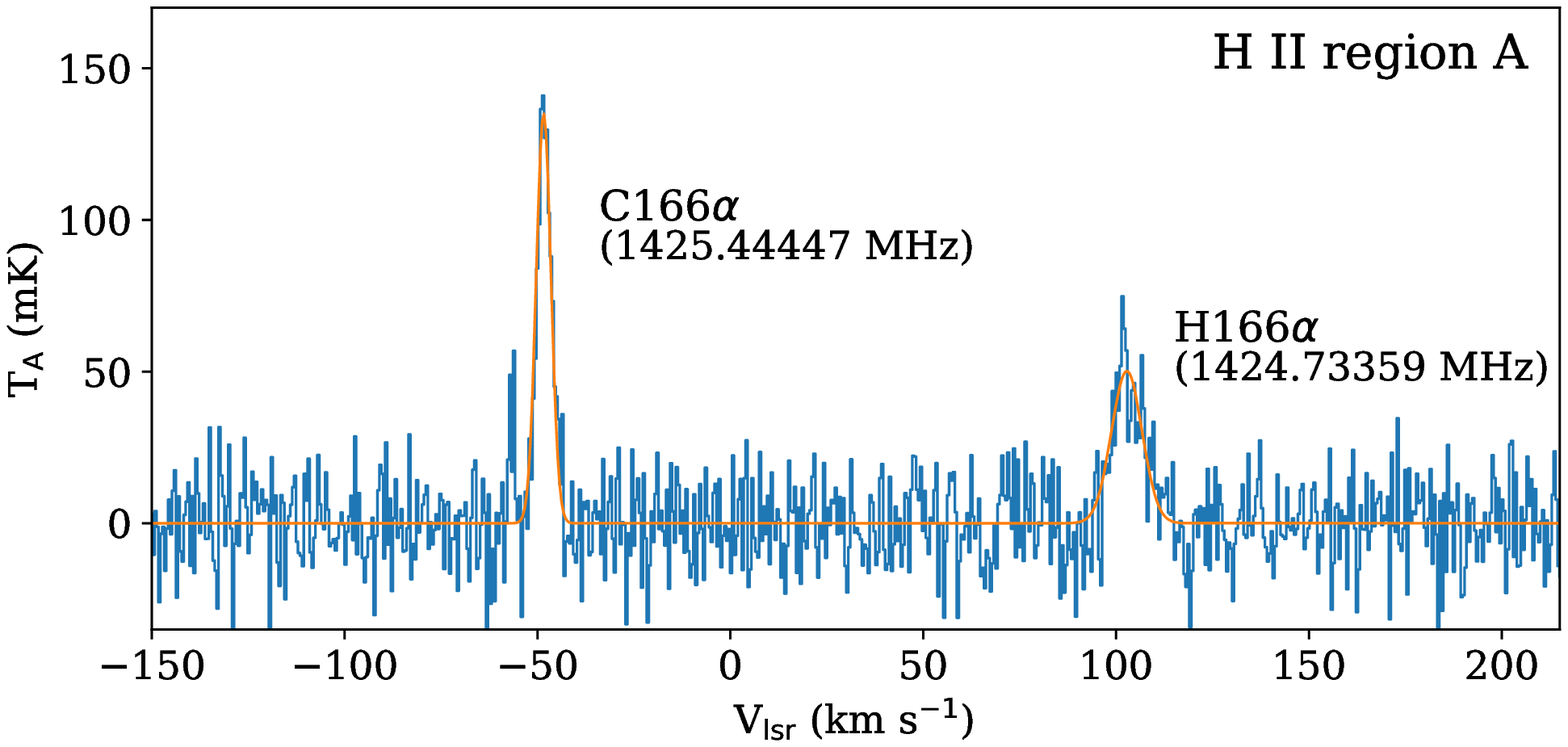}
    \includegraphics[width=0.45\textwidth]{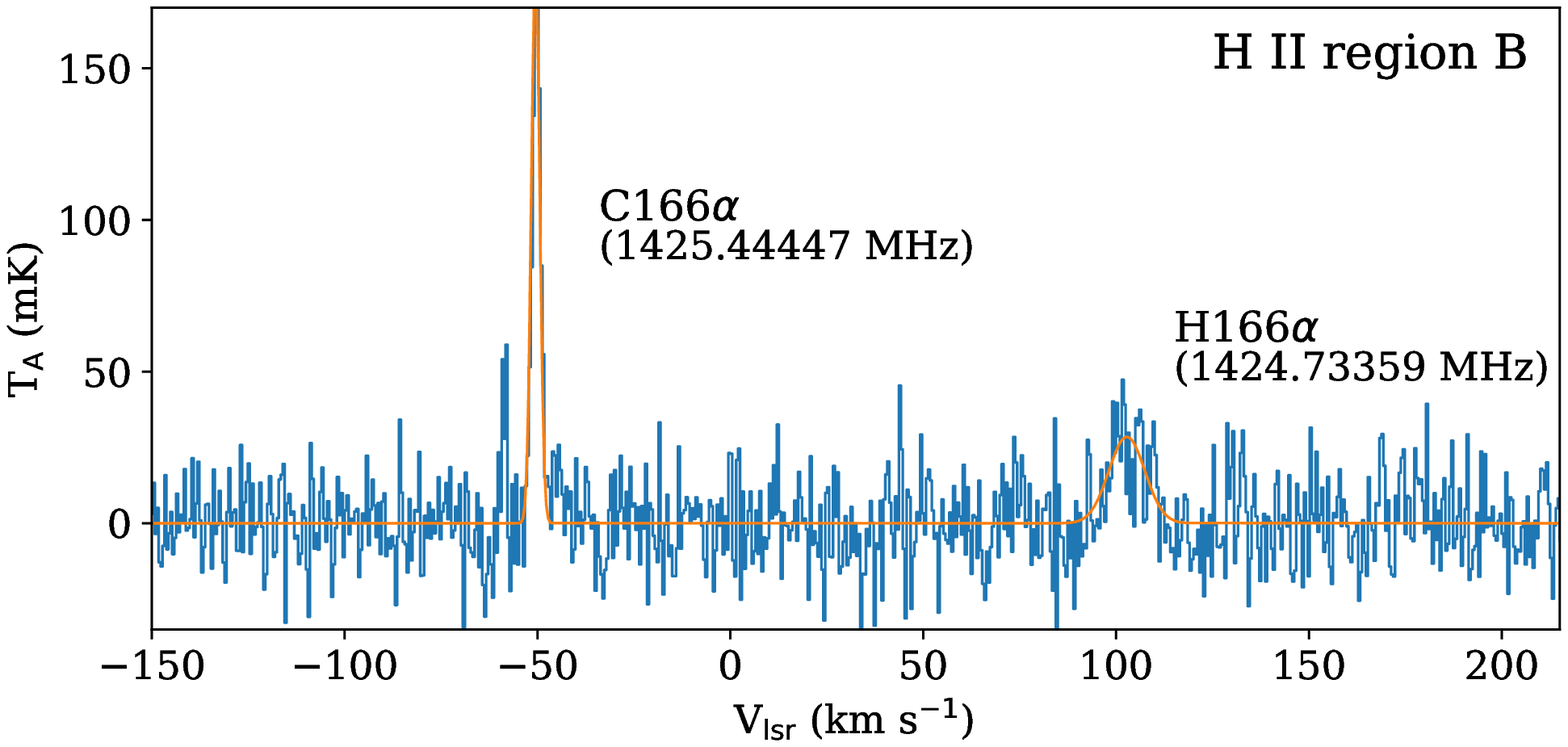}
    \includegraphics[width=0.45\textwidth]{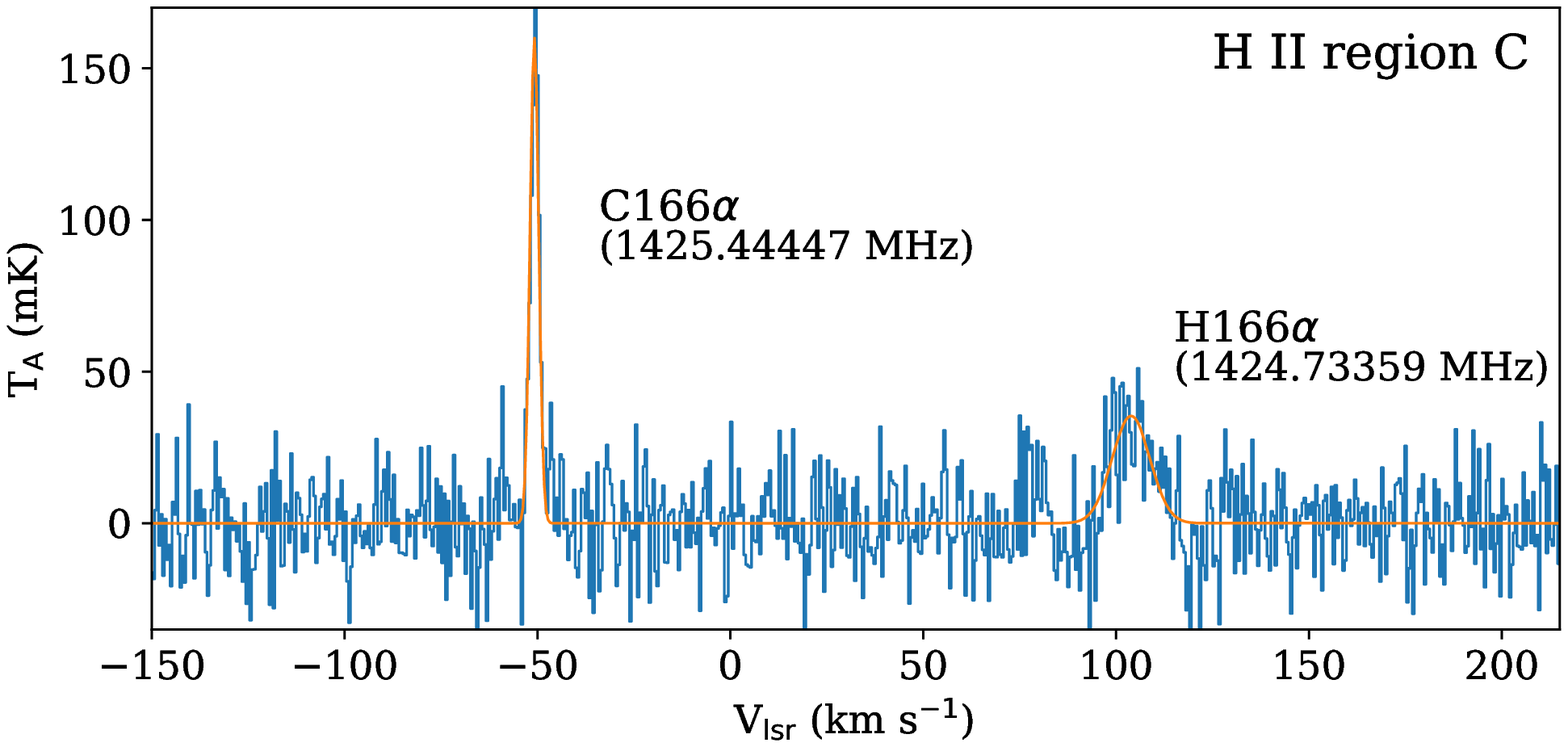}
    \includegraphics[width=0.45\textwidth]{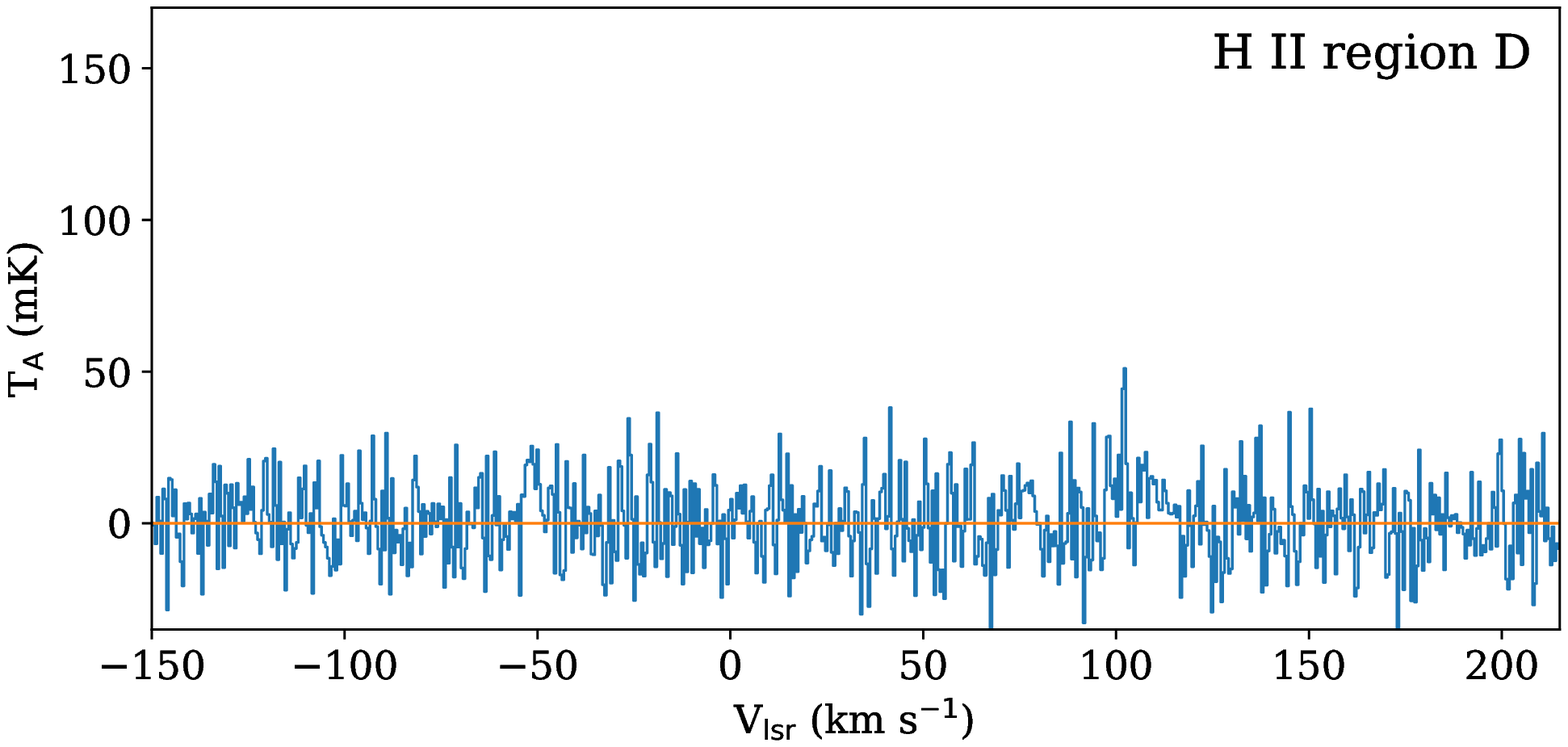}
    \vspace{-2mm}
    \caption{ The C166$\alpha$ and H166$\alpha$ RRLs of four different HII regions from the FAST observation. The orange smooth curves is a double-Gaussian fit to the observed spectra.
    }
    \label{fig:hii}
\end{figure}

\begin{table*}
\renewcommand\arraystretch{0.9}
\begin{center}
\tabcolsep 1.0mm\caption{Basic and fitted parameters of C166$\alpha$ and H166$\alpha$ spectra for four \HII regions}
\def\temptablewidth{1\textwidth}%
\begin{tabular}{llccccccccccccr}
\hline\hline\noalign{\smallskip}
 &&      &&&&& C166$\alpha$   &  &  & H166$\alpha$ &  \\
\cline{6-8}\cline{9-11}
Name  & RA & Dec &  Flux  & $T_{\rm mb}$  &$V_{\rm LSR}$ &FWHM   & $T_{\rm mb}$  &$V_{\rm LSR}$ &FWHM & RMS & $N_{\rm Ly}$  & $t_{\rm HII}$  \\
  &(hms) & (dms) & (Jy)  & (mK)   & (km s$^{-1}$)   &  (km s$^{-1}$)     &  (mK)  & (km s$^{-1}$)  & (km s$^{-1}$) & (mK) & (ph s$^{-1}$) & ($\times$10$^{5}$ yr)\\   
    \hline\noalign{\smallskip} 
\HII A & 02:28:21.9 & 61:28:34.8 & 19.6(0.3)& 135.2(6.4) & -48.4(0.1) & 2.7(0.1) & 50.2(4.4) & -46.7(0.4) & 5.6(0.6) & 17.7 & 5.91$\times$10$^{48}$ & 2.4\\
\HII  B  & 02:28:17.4 & 61:32:12.2 & 19.3(0.1) & 186.5(8.6) & -50.5(0.1) & 1.5(0.1) & 28.5(4.1) & -46.7(0.8) & 6.6(1.1) & 14.6 &5.82$\times$10$^{48}$ & 3.5\\
\HII  C & 02:28:32.3 & 61:30:51.1 & 19.6(0.3) &160.0(8.4) & -50.8(0.1) & 1.6(0.1) & 35.4(4.1) & -45.6(0.7) & 6.9(0.9) & 16.7 &5.91$\times$10$^{48}$& 1.7\\
\HII  D & 02:29:02.5 & 61:32:21.6 & 19.3(0.1)& -- & -- & -- & -- & -- & -- & 13.3 & 5.82$\times$10$^{48}$ &1.7\\
\noalign{\smallskip}\hline
\end{tabular}\end{center}
{ Note: RA and DEC are the central equatorial coordinates of the four \HII regions. Flux is the observed specific flux density. $T_{\rm mb}$ is main-beam brightness temperature. $V_{\rm LSR}$ and FWHM are the velocity and full-width half maxium of the radio recombination line. The Mean RMS is reduced root mean square of the spectrum. $N_{\rm Ly}$ is the measured ionizing luminosity, and $t_{\rm HII}$ is the the obtained dynamical ages.}
\label{Table:dust-clump-2}
\end{table*}

\begin{figure}
    \centering
    \includegraphics[width=0.45\textwidth]{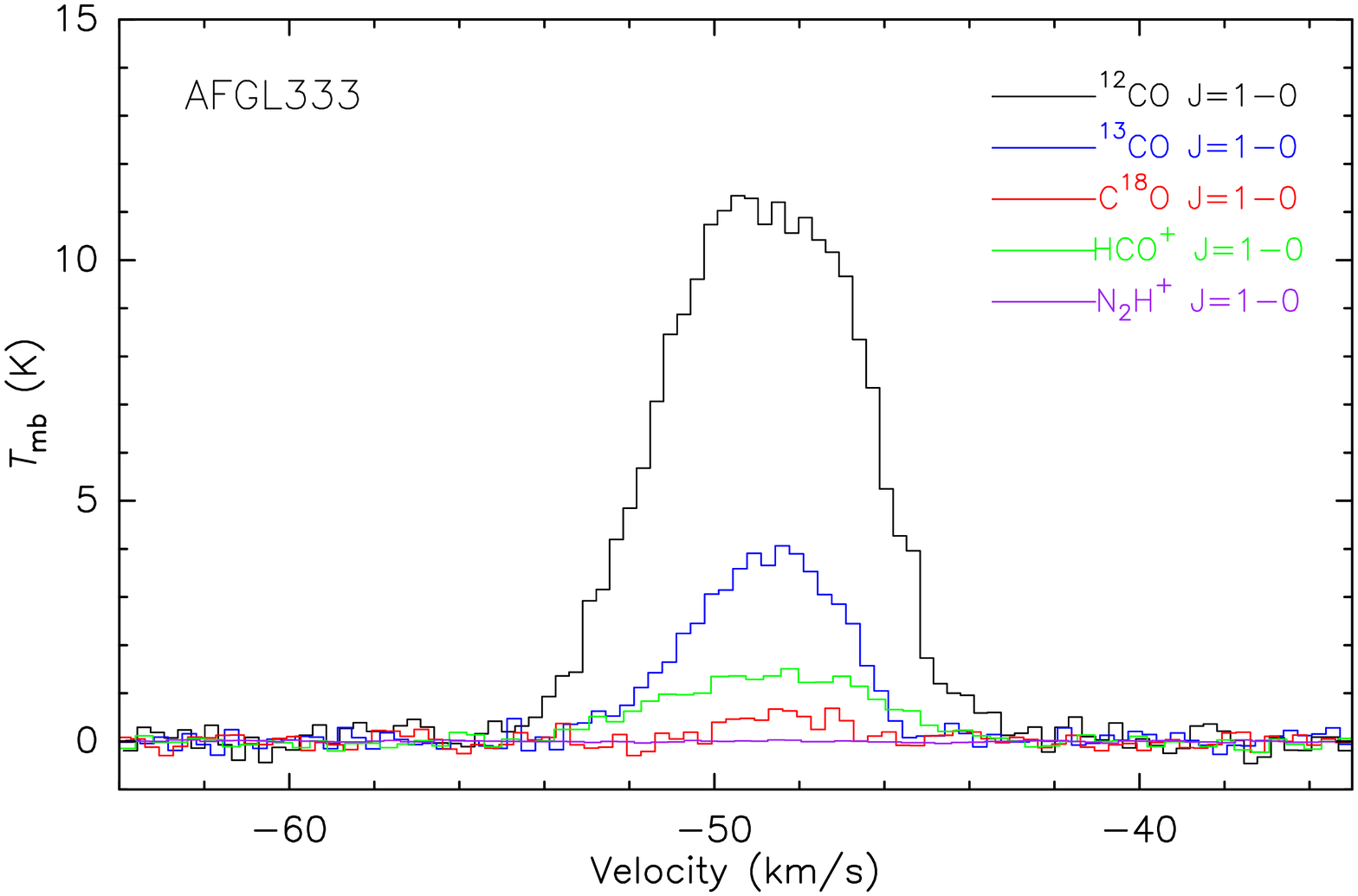}
    \caption{ 
    Averaged spectra of $^{12}$CO $J$=1-0, $^{13}$CO $J$=1-0, C$^{18}$O $J$=1-0, HCO$^{+}$ $J$=1-0, and N$_{2}$H$^{+}$ $J$ =1-0 of the AFGL 333-Ridge. For clarity, the intensity of HCO$^+$ $J$=1-0 is multiplied by factor 7.
    }
    \label{fig:lines}
\end{figure}

\subsection{Herschel column density and temperature  maps}
\label{herschel}

To study the influence of the \HII regions on the AFGL 333-Ridge, it is essential to estimate its gas density and dust temperature. We use Hi-GAL survey data  from 160 to 500 $\mu$m to construct a column density map ($N_{\rm H_{2}}$) of the hydrogen molecule. Because 70 $\mu$m emission is dominated by the warm dust heated by UV emission, we did not use the {\it{\it Herschel}} 70 $\mu$m data. Using the {\it Getsources} algorithm  \citep{2010A&A...518L.103M,2012A&A...542A..81M,2013A&A...560A..63M}, the dust temperature and column density maps of the AFGL 333-Ridge were created by  fitting the spectral energy distributions (SEDs) on a pixel-by-pixel basis \citep{2013A&A...550A..38P}. 

Figure~\ref{fig:sed} shows the column density and dust temperature map overlaid with the 1.1 mm continuum emission and the 21 cm radio emission. For the AFGL 333-Ridge traced by the 1.1 mm continuum emission in Fig.~\ref{fig:sed}, we obtain that the mean column density is 4.0$\times$10$^{22}$ cm$^{-2}$, and the mean dust temperature is 16 K. A noticeable feature in these two maps that the AFGL 333-Ridge exhibits higher column density, but lower temperature. \cite{2013ApJ...766...85R} also found a clear anti-correlation between column density and dust temperature in the most part of W3 GMC. In Figure~\ref{fig:sed}, the column density of some regions is higher than 1.5$\times$10$^{23}$ cm$^{-2}$. \cite{2013ApJ...766...85R} referred to $N(\rm H_{2})\sim1.8\times10^{23}$ cm$^{-2}$ as a massive star formation threshold according to the prediction of \cite{2008Natur.451.1082K}. Moreover, the regions adjacent to the four \HII regions have higher temperatures, which is about 23 K based on Figure~\ref{herschel}(right). The \HII region D shows a ring-like structure with an opening towards the southwest in the temperature map.

\begin{figure*}
    \centering
    \includegraphics[width=0.9\textwidth]{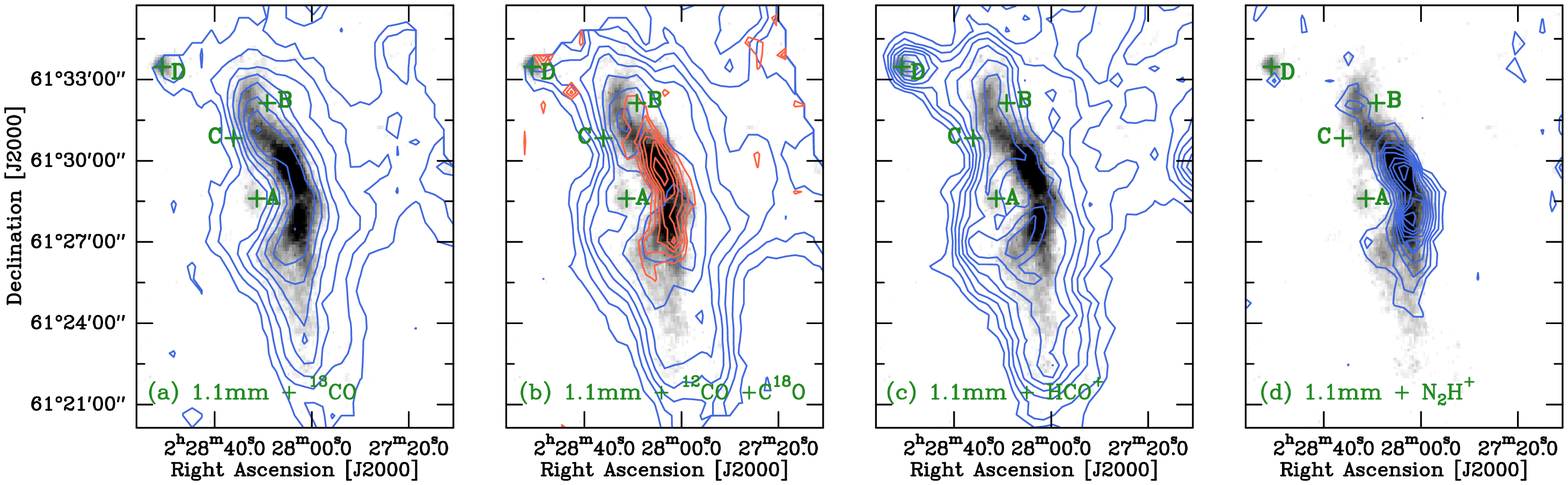}
    \caption{The background is the BGPS 1.1mm emission map, and the green plus symbols indicate the central position of four \HII regions. (a)The blue contours are the $^{13}$CO J =1-0 intensity emission which start from 6 K km s$^{-1}$ (3$\sigma$) in steps of 3$\sigma$. (b)$^{12}$CO $J$=1-0 intensity contours in blue start from 20 K km s$^{-1}$ (4$\sigma$) in steps of 3$\sigma$ and C$^{18}$O $J$=1-0 intensity contours in red start from 4.5 K km s$^{-1}$ (5$\sigma$) in steps of 1.5$\sigma$. (c)HCO$^{+}$ $J$=1-0 intensity contours start from 0.6 K km s$^{-1}$ (4$\sigma$) in steps of 2$\sigma$. (d)N$_{2}$H$^{+}$ $J$=1-0 intensity contours start from 0.09 K km s$^{-1}$ (3$\sigma$) in steps in 6$\sigma$.
     }
    \label{fig:four}
\end{figure*}

\begin{figure*}
    \centering
    \includegraphics[width=0.85\textwidth]{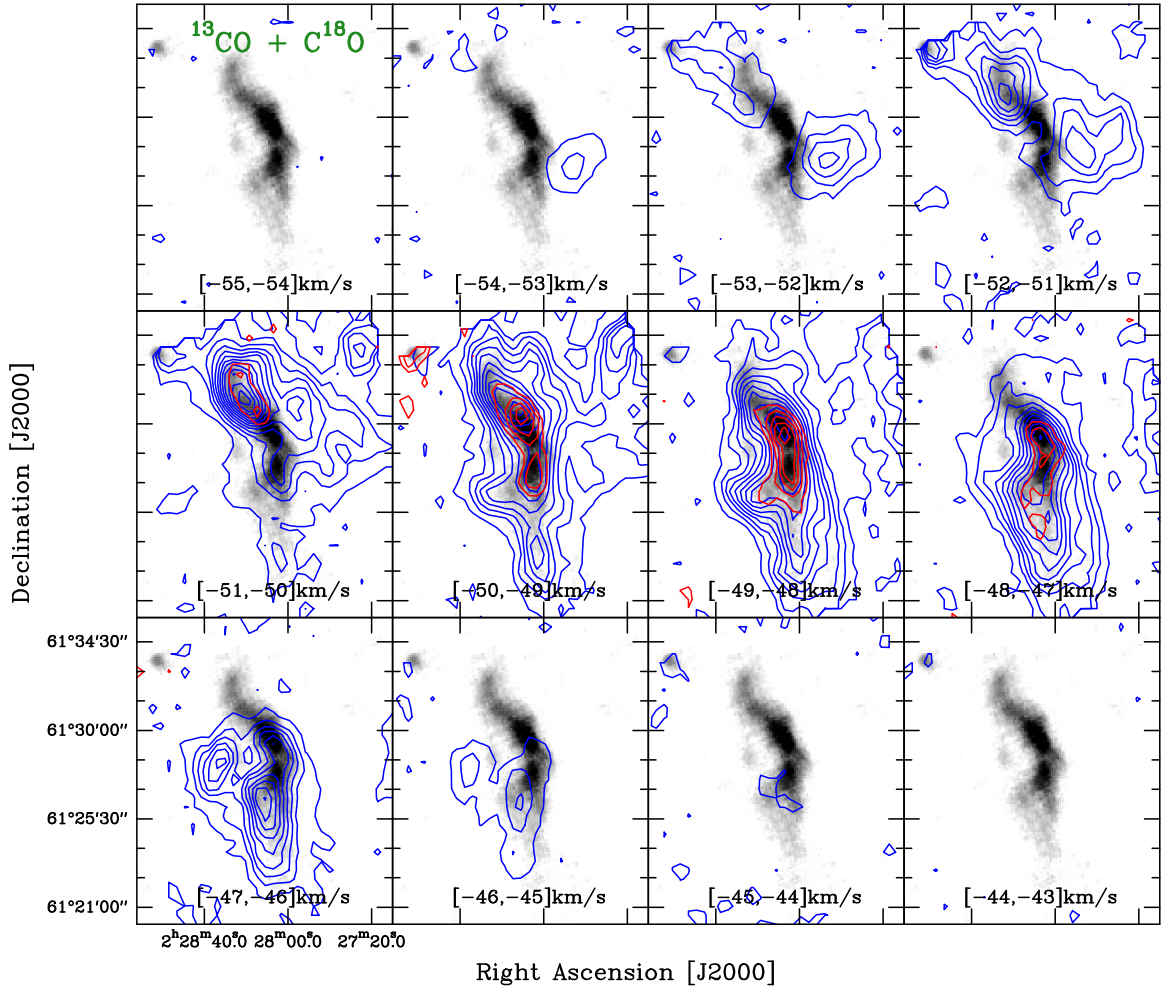}
    \caption{ 
     $^{13}$CO $J$=1-0 velocity channel maps (in blue contours) and C$^{18}$O $J$=1-0 velocity channel maps (in red contours) overlaid on the 1.1 mm emission (gray) in the steps of 1 km s$^{-1}$. The $^{13}$CO intensity contours start from 2.4 K km s$^{-1}$ (3$\sigma$) in steps of 3$\sigma$, and the C$^{18}$O intensity contours stared with  the levels of 1.8 K km s$^{-1}$ (6$\sigma$) in steps of 3$\sigma$. The molecular emission is integrated over a velocity interval, which is given in each panel.
    }
    \label{fig:13co}
\end{figure*}

\begin{figure*}[h]
    \centering
    \includegraphics[width=0.85\textwidth]{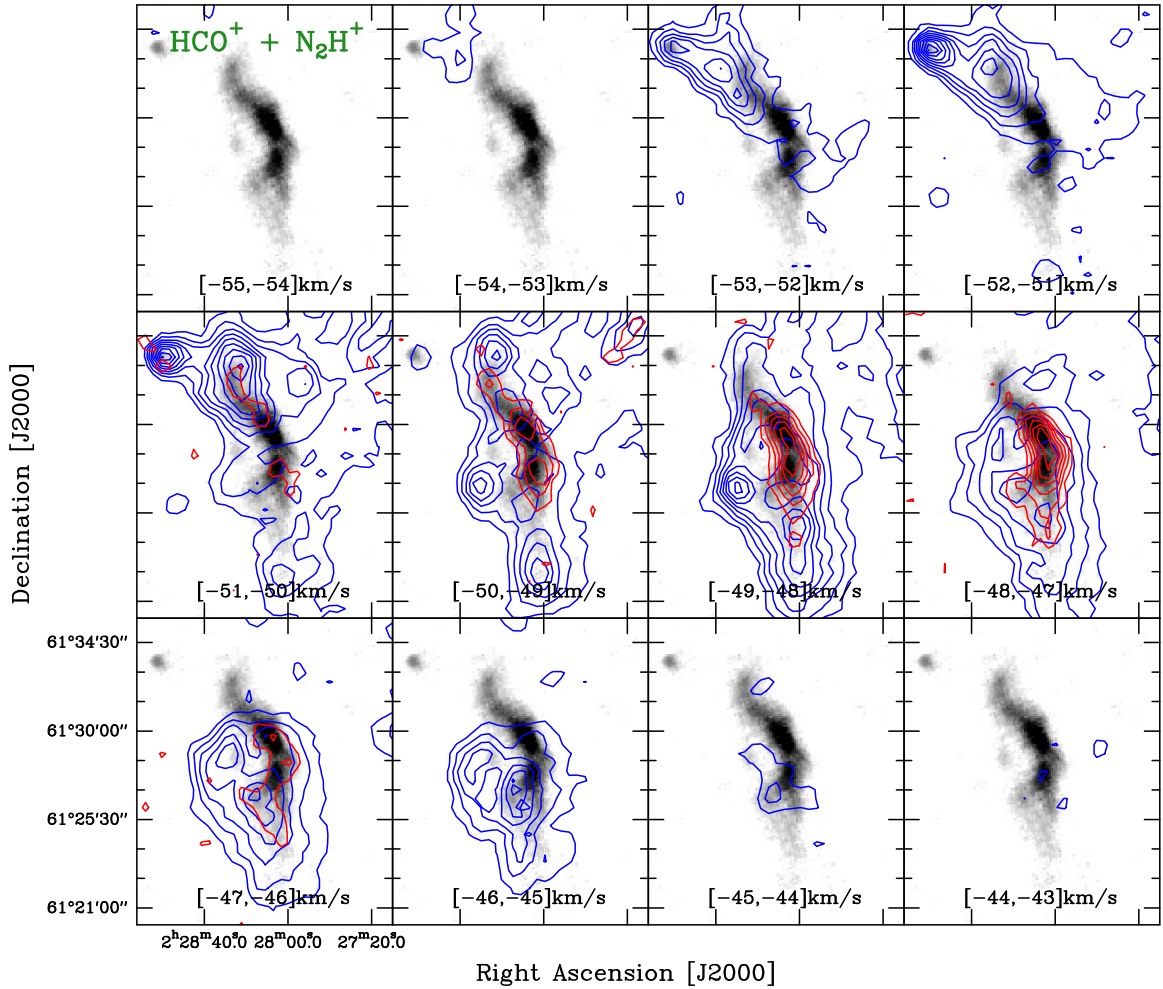}
    \caption{ 
     HCO$^{+}$ $J$=1-0 channel maps (blue contours) and N$_{2}$H$^{+}$ $J$=1-0 channel maps (red contours) from the integrated emission in the steps of 1 km s$^{-1}$ overlaid on the BGPS 1.1 mm emission. The blue contours start from 0.25 K km s$^{-1}$ (5$\sigma$) in steps of 3$\sigma$. The red contours start from 0.075 K km s$^{-1}$ (3$\sigma$) in steps of 3$\sigma$. The molecular emission is integrated over a velocity interval, which is given in each panel.}
    \label{fig:hco}
\end{figure*}

\begin{figure*}
    \centering
    \includegraphics[width=0.9\textwidth]{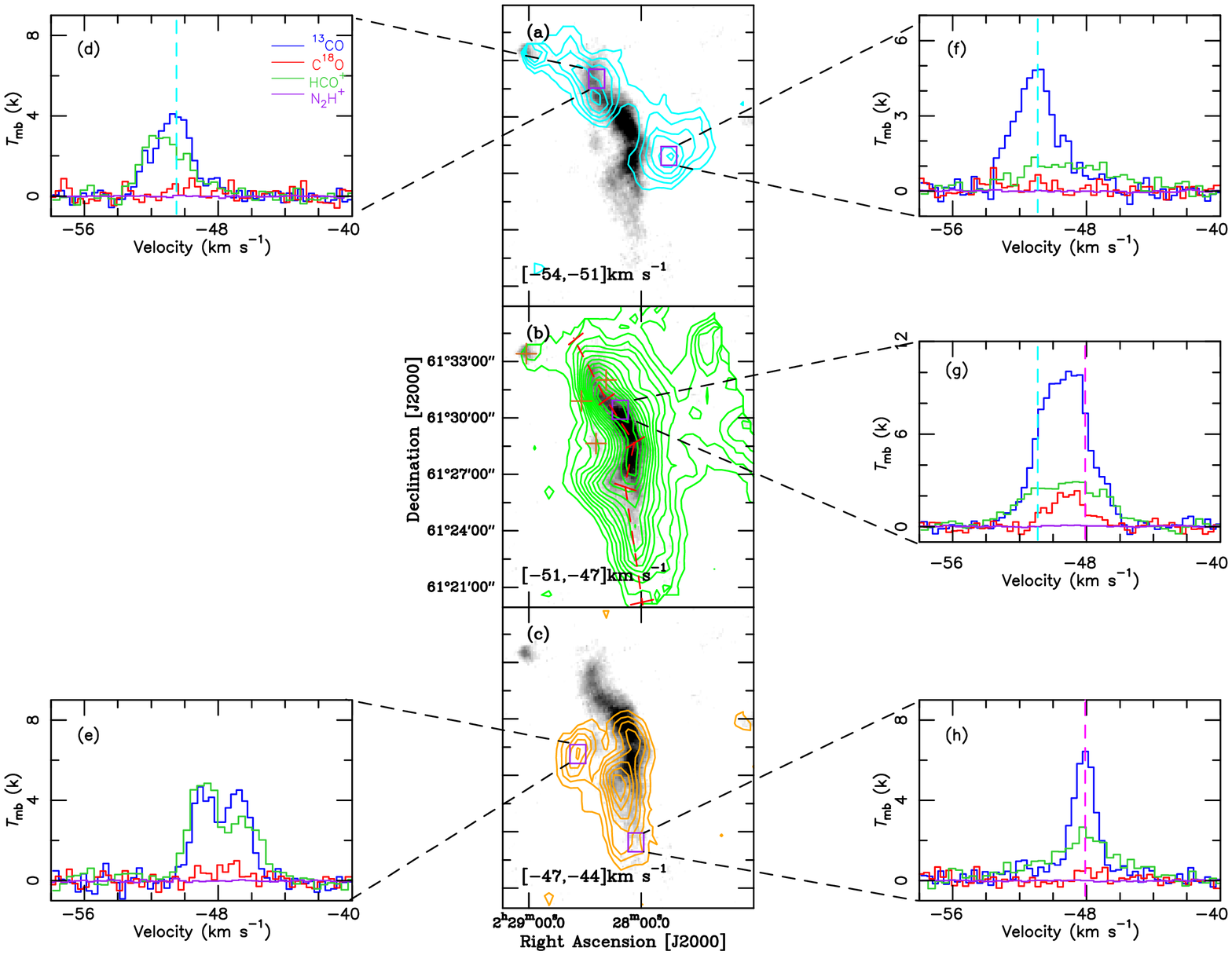}
    \caption{(a)-(c) panels. The integrated-intensity maps of $^{13}$CO $J$=1-0 overlaid on the 1.1 mm emission. The velocity intervals are shown in each panel. The four sections along the red dashed line  in (b)  panel indicate the direction of position-velocity diagram (see Figure~\ref{fig:pv}).  The brown cross symbols indicate the central position of four H II regions. The region highlighted by purple solid lines is used for extracting the molecular lines. The $^{13}$CO intensity contours start from  3 K km s$^{-1}$  (3$\sigma$) in steps of 2$\sigma$. (d)-(h) panels. Averaged spectra of $^{13}$CO $J$=1-0, C$^{18}$O $J$=1-0, HCO$^+$ $J$=1-0 and N$_2$H$^+$ $J$=1-0 over the area  marked by purple rectangles in (a)-(c) panels. The cyan and magenta dashed lines show the peak velocities of the two different components, respectively.
    }
    \label{fig:spectra}
\end{figure*}

\subsection{The RRL Spectra of Four \HII Regions}
\label{sect:HII} 
In order to determine whether the \HII regions interacted with the AFGL 333-Ridge, we use the FAST to observe \HII regions A, B, C, and D and use the double-Gaussians to fit the observed spectra. Figure~\ref{fig:hii} shows the C166$\alpha$ and H166$\alpha$ RRLs of the four \HII regions. However, we did not detect the emission of the radio recombination lines in the mean RMS of 13.3 mK towards the \HII region D. In each panel of Figure~\ref{fig:hii}, we compare the H166$\alpha$ RRL with the C166$\alpha$ RRL, the C166$\alpha$ spectrum with a narrow line width is more suitable to trace the RRL velocity of ionized gas \citep{2020ApJ...893L...5X}. Table~\ref{Table:dust-clump-2} gives the basic and fitted parameters of C166$\alpha$ and H166$\alpha$ spectra for these four \HII regions. From Table~\ref{Table:dust-clump-2}, we obtain that the  RRL velocity of \HII region A is $-$48.4 \kms, while \HII regions B and C have approximately the same RRL velocity. Moreover, the 1.4 GHz radio continuum emission of \HII region B is connected with that of \HII region C in space. Hence, we adopt $-$50.7 \kms as the mean RRL velocity of \HII regions B and C system.

\subsection{Molecular Gas Distribution}
\label{sect:molecular}

Figure~\ref{fig:lines} shows the averaged spectra of $^{12}$CO $J$ =1-0, $^{13}$CO $J$=1-0, C$^{18}$O $J$=1-0, HCO$^+$ $J$=1-0, and N$_2$H$^+$ $J$=1-0 over the AFGL 333-Ridge region, which is indicated by an orange box in Figure~\ref{fig:sed}. The peaks of $^{12}$CO, $^{13}$CO, C$^{18}$O, and HCO$^+$ spectra are near $-$49.0 $\pm$ 0.5 km s$^{-1}$ with a velocity range from $-$55 to $-$43 km s$^{-1}$. For the N$_2$H$^+$ $J$=1-0 spectrum, we cannot see a prominent temperature peak. Using the above velocity range, we made the integrated-intensity maps of five different spectra superimposed on the BGPS 1.1 mm emissions, shown in Figure~\ref{fig:four}. All five molecular emission of the AFGL 333-Ridge in Figure~\ref{fig:four} show a filamentary morphology extended from the north to the south. Moreover, both the C$^{18}$O and N$_2$H$^+$ emission, tracing the densest region of  the AFGL 333-Ridge, are associated well with the BGPS 1.1 mm emission.

Based on the velocity range of $-$55 to $-$43 km s$^{-1}$, we make channel maps of $^{13}$CO $J$=1-0, C$^{18}$O $J$=1-0, HCO$^+$ $J$=1-0, and N$_2$H$^+$ $J$=1-0. Figure~\ref{fig:13co} shows the $^{13}$CO $J$=1-0 channel maps (blue contours), and C$^{18}$O $J$=1-0 channel maps (red contours) overlaid on the 1.1 mm emission (gray scale), whose velocity varies from $-$54 to $-$44 km s$^{-1}$ in steps of 1 km s$^{-1}$.  The $^{13}$CO $J$=1-0 emission of the AFGL 333-Ridge has two velocity components (see Figure~\ref{fig:13co}). Component 1 is located in velocity range from $-$54 to $-$47 km s$^{-1}$, while component 2 from $-$51 to $-$44 km s$^{-1}$. It suggests that the two components overlap with each other in velocity, ranging from $-$51 to $-$47 km s$^{-1}$. Compared with the $^{13}$CO $J$=1-0 emission, the C$^{18}$O $J$=1-0 emission traces the denser region of the AFGL 333-Ridge, whose velocity varies from $-$51 to $-$47 km s$^{-1}$. We made the HCO$^+$ $J$=1-0 and N$_2$H$^+$ $J$=1-0 channel maps to further study this phenomenon. In Figure~\ref{fig:hco}, the channel maps of HCO$^+$ emission also show two different velocity components more clearly, which is the same as the $^{13}$CO $J$=1-0 channel maps. From $-$53 km s$^{-1}$ to $-$50 km s$^{-1}$, the HCO$^+$ $J$=1-0 emission shows a compact clump structure in the north-east, which is associated with IRAS 02252+6120 and \HII region D. The N$_2$H$^+$ $J$=1-0 emission in Figure~\ref{fig:hco} also traces the dense region of the AFGL 333-Ridge, whose velocity varies from $-$51 to $-$47 km s$^{-1}$. The velocity range of the C$^{18}$O $J$=1-0 and N$_2$H$^+$ emission is almost the same as that of the overlapping part in the $^{13}$CO $J$=1-0 and HCO$^+$ $J$=1-0 emission.

In order to show the positional correlation of the two velocity components mentioned above, we make the integrated-intensity maps of $^{13}$CO $J$=1-0, as shown in the panels of Figure~\ref{fig:spectra}. The integrated-velocity ranges are shown in each panel. In Figure~\ref{fig:spectra}(b), the CO component of the overlap velocity ($-$51 to $-$47 km s$^{-1}$) shows an elongated structure extended from north to south, which is associated well with the 1.1 mm dust emission. Figure~\ref{fig:spectra}(a), shows that component 1 of $-$54 to $-$51 \kms presents a northeast-southwest filamentary structure, which is coincident with the northern section of the AFGL 333-Ridge.  While component 2 of $-$47 to $-$44 \kms in Figure~\ref{fig:spectra}(c) also shows a northeast-southwest filamentary structure, which is mainly associated with the southern section of the AFGL 333-Ridge. The average spectra of $^{13}$CO $J$=1-0, C$^{18}$O $J$=1-0, HCO$^+$ $J$=1-0, and N$_2$H$^+$ $J$=1-0 over the five purple rectangles are shown in Figures~\ref{fig:spectra}(d)-(h). In Figures~\ref{fig:spectra}(d) and \ref{fig:spectra}(f), the $^{13}$CO $J$=1-0 spectra show a peak at $-$50.5 km s$^{-1}$ highlighted by the cyan dashed line, and spectra in Figure~\ref{fig:spectra}(h) show a peak at $-$48.0 km s$^{-1}$ highlighted by a magenta dashed line. These two peaks may indicate this molecular cloud has two different velocity components. The flat-top between the two peaks ranges from $-$50.5 km s$^{-1}$ to $-$48.0 km s$^{-1}$, as shown in Figure~\ref{fig:spectra}(g). We also check the spectra in \HII region A position, as shown in Figure~\ref{fig:spectra}(e). Both $^{13}$CO $J$=1-0 and HCO$^+$ $J$=1-0 spectra show a double-peak profile. Because the optical thin spectral line C$^{18}$O $J$=1-0 is relatively weak, we cannot judge whether the double peak is composed of two components or the absorption of a single component.

 \begin{figure*}
    \centering
    \includegraphics[width=0.46\textwidth]{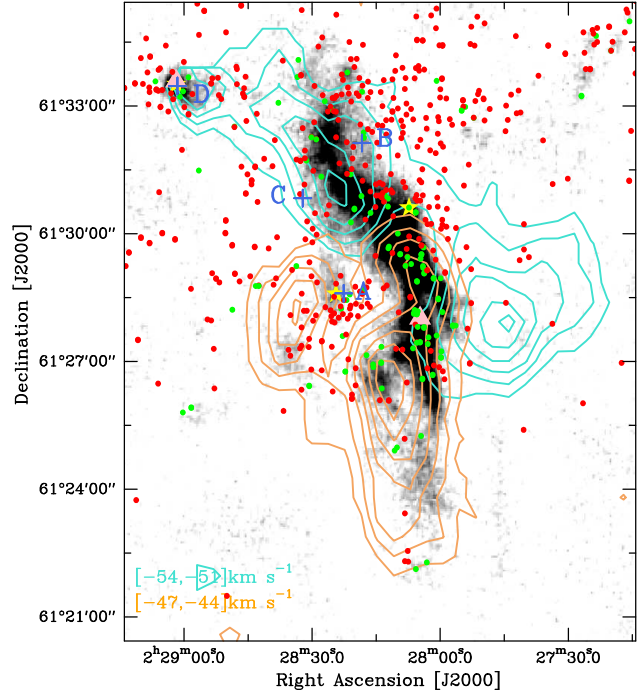}
    \caption{ Overlay of the $^{13}$CO emission contours on the 1.1mm emission in grayscale. The cyan contours show the $^{13}$CO emission with a integrated-velocity range of $-$54 km s$^{-1}$ to $-$51 km s$^{-1}$, while the orange contours represent the $^{13}$CO emission at a integrated-velocity range of $-$47 km s$^{-1}$ to $-$44 km s$^{-1}$. The green dots and red dots represent the identified Class I and Class II YSOs selected from the candidate sources in this region by \cite{2011ApJ...743...39R} and \cite{2016ApJ...822...49J}.  The blue pluses illustrate the locations of four H II regions. The three IRAS sources labeled as yellow star symbols, and two pink triangle symbols indicate H$_2$O masers.}
    \label{fig:collison}
\end{figure*}

\subsection{Physical parameters}
\subsubsection{The column density and mass of the AFGL 333-Ridge derived from CO data}
\label{pp}
Assuming local thermal equilibrium \citep{2009tra..book.....W,2015PASP..127..299S}, we use the $^{13}$CO $J$=1-0 emission to determine the column density and mass of the AFGL 333-Ridge. While the optical thick $^{12}$CO $J$=1-0 emission is used to determine excitation temperature, which can be given by the equation $T_{\rm ex} = \frac{5.53}{{\rm ln}[1+5.53/(T_{\rm mb}+0.82)]}$ \citep{1991ApJ...374..540G}, where $T_{\rm mb}$ is the corrected main-beam brightness temperature of $^{12}$CO. We assume that the excitation temperature of $^{13}$CO equal to the excitation temperature of $^{12}$CO. The optical depth ($\tau$) is estimated via the following equation $\tau(^{13}{\rm CO}) = -{\rm ln}[1-\frac{T_{\rm mb}}{5.29/[{\rm exp}(5.29/T_{\rm ex})-1]-0.89}]$ \citep{1991ApJ...374..540G}. The column density of $^{13}$CO can be estimated by the following equation \citep{1991ApJ...374..540G}:
\begin{equation}\label{eq:3}
\begin{split}
N = 4.71\times10^{13}\frac{T_{\rm ex}+0.88}{{\rm exp}(-5.29/T_{\rm ex})}\times T\times
W {\rm cm}^{-2},
\end{split}
\end{equation}
where $T_{\rm ex}$ is the mean excitation temperature of molecular gas. $T$ equals $\frac{\tau}{1-{\rm exp}(-\tau)}$. $W$ is $\int T_{\rm mb} dv$ in units of K km s$^{-1}$, where $dv$ is the velocity range and  $T_{\rm mb}$ is the corrected main-beam temperature of $^{13}$CO $J$=1-0. Then the H$_2$ column density is calculated using $N({\rm H_2})$ = $X(^{13}{\rm CO})\times N$. Taking the conversion factor of $X(^{13}{\rm CO})$ $\sim$ 7 $\times$ 10$^5$ \citep{1994ARA&A..32..191W}, we derive the column density of 2.3$\times$10$^{22}$ cm$^{-2}$, 2.4$\times$10$^{22}$ cm$^{-2}$ and 2.6$\times$10$^{22}$ cm$^{-2}$ for the molecular structures at different velocity ranges in Figures \ref{fig:spectra}(a)-(c). The obtained mean column densities by CO data are roughly equal to that (4.0 $\times$10$^{22}$ cm$^{-2}$) estimated by the {\it {\it Herschel}} data in Sect~\ref{herschel}.
The mass of the AFGL 333-Ridge can be written as
\begin{equation}\label{eq:4}
      M_{\rm H_{2}} = S\mu_gm(\rm H_{2})\it N(\rm H_{2}),
\end{equation}
where $S$ is the area of the molecular cloud, which is measured through the $^{13}$CO emission contours in Figures~\ref{fig:spectra}(a)-(c). $\mu_g$ is the mean atomic weight (1.36) of the gas, and $m(\rm H_{2})$ is the mass of a hydrogen molecule. For the AFGL 333-Ridge, we take the distance of 2.0 kpc \citep{2006Sci...311...54X}. Using the equation~\ref{eq:4}, we derive that the mass of each component in Figures~\ref{fig:spectra}(a)-(c) are 1.0$\times$10$^{4}$ $M_{\odot}$, 1.6$\times$10$^{4}$ $M_{\odot}$, and 9.0$\times$10$^{3}$ $M_{\odot}$, respectively. 

\subsubsection{The physical properties of our \HII regions}
The four \HII regions are identified in our observed region. We can calculate the dynamical ages of the \HII regions. The ionizing luminosity was given by $N_{\rm Ly}=7.54\times10^{46}(\frac{\nu}{\rm GHz})^{0.1}(\frac{T_4}{\rm K})^{-0.45}(\frac{S_v}{\rm Jy})(\frac{d}{\rm kpc})^2\rm s^{-1}$ \citep{1992ARA&A..30..575C}, where $\nu$ is the frequency of the observation, $T_4$ is the effective electron temperature in units of 10$^4$ K, $S_v$ is the observed specific flux density, and $d$ is the distance to \HII regions. Here we adopt an effective electron temperature of 10$^4$ K, and take the distance of the AFGL 333-Ridge (2 kpc) as those of the four \HII regions. We measured that the flux density of \HII regions A, B, C and D are 19.6 Jy, 19.3 Jy, 19.6 Jy and 19.3 Jy at 1.4 GHz, respectively. The derived $N_{\rm Ly}$ for the four \HII regions are listed in Table~\ref{Table:dust-clump-2}. Assuming the expansion of \HII region in a homogeneous medium, the dynamical age is estimated by \cite{1980pim..book.....D}:
\begin{equation}\label{eq:7}
      t_{\rm HII} = \frac{4R_{\rm s}}{7c_{\rm s}}[(\frac{R_{\rm HII}}{R_{\rm s}})^\frac{7}{4}-1],
\end{equation}
where $R_{\rm s} = (\frac{3N_{\rm Ly}}{4\pi{n_i}^2{\alpha_b}})^{\frac{1}{3}}$ is the radius of  Str{\"{o}}mgren sphere. $n_{\rm i}$ is the initial volume density of gas, and $\alpha_B$ = 2.6$\times 10^{-13}$ cm$^3$ s$^{-1}$ is the hydrogen recombination coefficient to all levels above the  ground level. Because the H II regions are located close to the AFGL 333-Ridge, we use the average volume-density H$_2$ of the ridge to replace $n_{\rm i}$. The average volume-density of AFGL 333-Ridge is given by $n(\rm H_{2})$=$N(\rm H_{2})/\it L$, where $N(\rm H_{2})$ is the mean CO column density (2.4$\times$10 $^{22}$ cm$^{-2}$) and $L$ is the equivalent width of the AFGL 333-Ridge. According to the first contour of CO in Figure~\ref{fig:spectra}(b), we measured three different places to get the average width of the molecular cloud (0.9 pc), and take it as the equivalent width of the AFGL 333-Ridge. The obtained  $n_{\rm i}$ is 8.6$\times 10^3$ cm$^{-3}$. In Equation~\ref{eq:7}, $c_{\rm s}$ is the isothermal sound speed of the ionized gas, which is adopted by 10 km s$^{-1}$.  According to the area of the 1.4 GHz emission contours in Figure~\ref{fig:arc}, we measure the radius of 1.0 pc, 1.2 pc, 0.8 pc and 0.8 pc for \HII regions A, B, C, and D, respectively. Finally, the obtained dynamical ages are 2.4$\times$10$^5$ yr, 3.5$\times$10$^5$ yr, 1.7$\times$10$^5$ yr and 1.7$\times$10$^5$ yr for \HII regions A, B, C, and D, respectively.

\begin{figure*}
    \centering
    \includegraphics[width=0.75\textwidth]{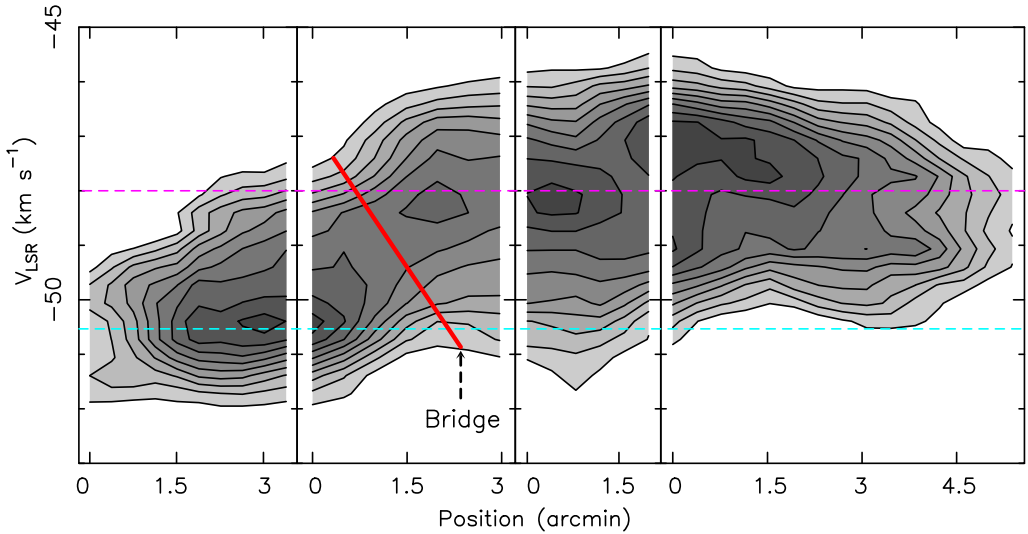}
    \caption{Position-velocity diagram of the $^{13}$CO $J$=1-0 emission along the different orientation of the AFGL 333-Ridge (see the red dashed line in Figure~\ref{fig:spectra} (b)). The cyan dashed line and the magenta dashed line indicate the two different LSR velocities.}
    \label{fig:pv}
\end{figure*}

\subsection{The distribution of YSOs}
\label{distribution}
Using  the Spitzer IRAC and MIPS observations, \cite{2016ApJ...822...49J} made deep JHKs photometry of the AFGL 333 region and identified many YSOs (young stellar objects), including low mass populations. To investigate the YSOs distribution and formation in the observed field, we use their candidate sources of Class I and Class II. Since the observation field of \cite{2016ApJ...822...49J} is not fully covered our CO observations, we complement the {\it{\it Spitzer}} YSOs catalog in this region by \cite{2011ApJ...743...39R}. From these two catalogs, we found 122 Class I YSOs and 464 Class II YSOs. Class I YSOs are protostars with circumstellar envelopes with a lifetimes of the order of $\sim10^{5}$ yr, while Class II YSOs are disk-dominated objects with a lifetimes of  $\sim10^{6}$ yr \citep{1994ApJ...420..837A}. Figure \ref{fig:collison} shows the spatial distribution map of these selected YSOs, overlaid on the $^{13}$CO $J$=1-0 emission of $-$54 km s$^{-1}$ to $-$51 km s$^{-1}$ (component 1) and $-$47 km s$^{-1}$ to $-$44 km s$^{-1}$ (component 2). In Figure~\ref{fig:collison}, the Class I YSOs are denoted by green dots, while red dots denote the Class II YSOs. It is clear that Class I YSOs accumulate along the dense ridge, where the two velocity components complement each other, shown in Figure~\ref{fig:collison}. In addition, \cite{2017PASJ...69...16N} has found that the AFGL 333-Ridge is associated with three IRAS sources and two H$_2$O masers which are well-known tracer of early star formation activity \citep{2007PASJ...59.1185S}.

\section{Discussion}
\label{sec:discussion}
Cloud-cloud collision rapidly produces a compressed layer, which can provide an essential condition for star formation \citep{2015ApJ...806....7T,2020PASJ..tmp..237F}. Hence, many researchers have proposed cloud-cloud collision  as an essential mechanism for star formation, even for the high-mass star, and star cluster formation \citep[e.g.,][]{2019ApJ...884...84D,2019A&A...631A...3M,2020ApJ...893L...5X,2020PASJ..tmp..237F}. The simulations and observations of the cloud-cloud collision have revealed that the colliding molecular clouds have three characteristic features: the spatially complementary distribution of the two colliding clouds, the bridge feature at the intermediate velocity range, and its flattened CO spectrum \citep[e.g.,][]{2010MNRAS.405.1431A,2014ApJ...792...63T,2015ApJ...806....7T,2017ApJ...835..142T,2018ApJ...859..166F}. The channel maps of $^{13}$CO $J$=1-0 and HCO$^+$ $J$=1-0 suggest that the AFGL 333-Ridge has two velocity components. Component 1 is located in velocity ranges from $-$54 to $-$47 km s$^{-1}$, while component 2 from $-$51 to $-$44 km s$^{-1}$.  The overlapping part of the velocity ($-$51 to $-$47 km s $^{-1}$) is the main component,  associated well with the 1.1 mm dust emission of the AFGL 333-Ridge.  Both the $^{13}$CO and HCO$^+$ spectra show a flattened profile (see Figure~\ref{fig:spectra} (g)). This scenario indicates that the two components are likely to collide with each other. To further examine the possibility of the collision, we made a position-velocity (PV) diagram (Figure~\ref{fig:pv}) along the red dashed line, which is highlighted in Figure~\ref{fig:spectra} (b). The PV diagram of $^{13}$CO presents a bridge feature, indicated by a red line in Figure~\ref{fig:pv}. Since the `bridge-feature' can be interpreted as the turbulent gas excited at the interface of the collision \citep{2017ApJ...835..142T}, it acts as a vital signature of two colliding clouds \citep{2015MNRAS.454.1634H}. These observed signatures illustrate that the two independent velocity components in the AFGL 333-Ridge have collided with each other. Also, the AFGL 333-Ridge shows a coherent structure (see Figure~\ref{fig:pv}). The identified bridge is almost as wide as the AFGL 333-Ridge. According to the Sect~\ref{pp}, the mass of the two collided components is 1.0$\times$10$^{4}$ $M_{\odot}$ and 9.0$\times$10$^{3}$ $M_{\odot}$. It suggests after passing through the collision, the two components with almost the same mass have gradually merged into a single structure. That is what we see as the whole AFGL 333-Ridge. 
 
The collision timescale of the two components can be determined by \cite{2013MNRAS.428.3425H} and \cite{2019ApJ...875..138D}:
\begin{equation}\label{eq:11}
   t_{\rm accum} = 2(\frac{l_{\rm fcs}}{0.5\,{\rm pc}})(\frac{{\nu}_{\rm rel}}{5\,{\rm km \ s}^{-1}})^{-1}(\frac{n_{\rm pstc}/n_{\rm prec}}{10}) \rm  Myr,
\end{equation}
where $l_{\rm fcs}$ is the collision length, ${\nu}_{\rm rel}$ is the relative velocity, $n_{\rm prec}$ and $n_{\rm pstc}$ is the mean density of pre-collision region, and post-collision region, respectively. The velocity separation of the two clouds is about 2.5 km s$^{-1}$. Nevertheless, it is difficult to observe the exact viewing angle of the collision. We assume the angle of the collision to the line of the sight is 45$^{\circ}$. Measuring the complementary part of the two velocity components in Figure~\ref{fig:collison}, the mean collision length scale can be calculated to be 3.5 pc (2.5 pc/$\sin45^{\circ}$), and the observed relative velocity is 3.5 km s$^{-1}$ (2.5 km s$^{-1}$/$\cos45^{\circ}$). However, the exact mean density of pre-collision and post-collision is hard to figure out. We assume that the density of the interaction area of AFGL 333-Ridge is the mean density of post-collision and the density of the surrounding area is pre-collision. Then we estimate the ratio between the pre- and post-collision densities which is about 1.2$\pm$0.2. Using this range of the  density ratios, we obtained the collision timescale of 2.4$\pm$0.4 Myr.  

In the observed region, we identified four \HII regions by using the 1.4 GHz continuum data (see Figure~\ref{fig:w34}). However, only \HII regions A, B, and C  are adjacent to the AFGL 333-Ridge on the 2D space projection. The RRL velocity ($-$48.4 km s$^{-1}$) of \HII region A is consistent with the peak velocity of component 2, while \HII region B and C ($-$50.7 km s$^{-1}$) is associated with component 1. The 1.1 mm dust and $^{13}$CO $J$=1-0 emission of the AFGL 333-Ridge shows a bow-like structure around \HII regions A and B.  Especially around \HII region B, the 4.5 $\mu$m emission shows an arc-like structure around it (see Figure~\ref{fig:arc}). The bow-like structures are likely to be formed by the  expansion of the \HII regions. For the dust temperature distribution of the AFGL 333-Ridge (see in Sect~\ref{herschel}), we found that the dust temperature around the \HII regions was relatively high. The area with higher dust temperature may be heated by the \HII regions. \cite{2017PASJ...69...16N} have confirmed that \HII region A interacts with the AFGL 333-Ridge. From the above observational phenomena, we conclude that \HII regions A, B, and C  may indeed be interacting with the AFGL 333-Ridge. Comparing the typical collision timescale in the AFGL 333-Ridge with the dynamical ages of \HII regions, the collision timescale is about an order of magnitude larger than the latter. Hence, we suggest that the interaction between the \HII regions and AFGL 333-Ridge is unlikely to drive the two clouds to merge. The age of the giant \HII region W4 associated with IC 1795 OB association was estimated to be 3--5 Myr by \cite{2005AJ....129..393O} and \cite{2011ApJ...733..113R}. The AFGL 333-Ridge is located toward the north-west of W4 in projection, see in Figure~\ref{fig:w34}. W4 is excited by a group of O stars located at the central part of the \HII region \citep{1997A&A...324..249L}. Both component 1 and component 2 show a northeast-southwest structure.  These two components are both elongated in the NE-SW direction, which seem to lie perpendicular to the expansion direction of W4. On a large scale (see Figure\ref{fig:w34}), in the other direction, we also find no other dynamic sources besides the AFGL 333-Ridge in the W3 GMC. Hence, the expansion of W4 may drive the two clouds to collide, then eventually merge to form the AFGL 333-Ridge.  In addition, the majority of the Class I YSOs gather in the interaction region of component 1 and component 2 (see Figure~\ref{fig:collison}). Comparing the typical collision timescales with the average ages of Class I and Class II YSOs, we suggest that the cloud-cloud collision has triggered the formation of YSOs in the AFGL 333-Ridge.

\section{Conclusions}
\label{sec:conclusions}
We performed multi-wavelength observations toward filamentary AFGL 333-Ridge. The major results of the present work are given below:

1. The molecular line data shows that the AFGL 333-Ridge constitutes two components, one at $-$54.0 to $-$47.0 km s$^{-1}$ and the other at $-$51.0 to $-$44.0 km s$^{-1}$. The mass of component 1 is 1.0$\times$10$^4$ M$_{\odot}$, while component 2 is 9.0$\times$10$^3$ M$_{\odot}$. Based on the $^{13}$CO position-velocity diagram, the two independent velocity components are interconnected in space by a bridge feature. The structure of the AFGL 333-Ridge is coherent in the PV diagram, which indicates that the two velocity components have collided and merged into a whole molecular cloud. The majority of the Class I YSOs are distributed within the collision region from two different components. From the comparison of the collision timescales and the average age of YSOs, we conclude that the cloud-cloud collision may create the YSOs formation in the AFGL 333-Ridge.

2.  The AFGL 333-Ridge is associated with three \HII regions. The gas adjacent to the \HII regions has a higher temperature ($\sim$23 K). By comparing the dynamical ages ((1.7--3.5)$\times$10$^5$ yrs) of  the \HII regions with the collision timescale (2.4$\pm$0.4 Myr) in the AFGL 333-Ridge, we conclude that although the \HII regions are interacting with the AFGL 333-Ridge, the influence of the \HII regions may not drive the two clouds to merge.  These two components are both elongated in the NE-SW direction, which seem to lie perpendicular to the expansion direction of W4. The expansion of W4 may drive the two clouds to collide, then eventually merge to form the AFGL 333-Ridge.

\acknowledgments
We thank the referee for insightful comments that improved the clarity of this manuscript. This work was supported by the Youth Innovation Promotion Association of CAS, and the National Natural Science Foundation of China (Grant Nos. 11873019 and 11933011).

\bibliographystyle{aasjournal}
\bibliography{ref}

\end{document}